\documentclass[nopacs,twocolumn,aps,pre,superscriptaddress,
nofootinbib, 
letterpaper]{revtex4-1}
\usepackage{graphicx}
\usepackage[caption=false]{subfig}
\usepackage{amsmath}
\usepackage{color}
\usepackage{placeins}
\usepackage{array}
\usepackage{siunitx}
\usepackage[utf8]{inputenc}
\usepackage{hyperref}

\newcommand*\chem[1]{\ensuremath{\mathrm{#1}}} 
\begin{document}
\onecolumngrid \noindent
This article may be downloaded for personal use only. Any other use requires prior permission of the author and AIP Publishing. This article appeared in the Journal of Chemical Physics (Vol.\ 151, Issue 11) and may be found at \url{https://doi.org/10.1063/1.5118875}. \medskip
\twocolumngrid

\title{
Magnetostriction in magnetic gels and elastomers as a function of the internal structure and particle distribution
}

\author{Lukas Fischer}
\email{lfischer@thphy.uni-duesseldorf.de}
\affiliation{Institut f{\"u}r Theoretische Physik II: Weiche Materie, 
Heinrich-Heine-Universit{\"a}t D{\"u}sseldorf, Universit{\"a}tsstra{\ss}e 1, D-40225 D{\"u}sseldorf, Germany}
\author{Andreas M. Menzel}
\email{menzel@thphy.uni-duesseldorf.de}
\affiliation{Institut f{\"u}r Theoretische Physik II: Weiche Materie, 
	Heinrich-Heine-Universit{\"a}t D{\"u}sseldorf, Universit{\"a}tsstra{\ss}e 1, D-40225 D{\"u}sseldorf, Germany}

\date{\today}

\begin{abstract}
Magnetic gels and elastomers are promising candidates to construct reversibly excitable soft actuators, triggered from outside by magnetic fields. These magnetic fields induce or alter the magnetic interactions between discrete rigid particles embedded in a soft elastic polymeric matrix, leading to overall deformations. It is a major challenge in theory to correctly predict from the discrete particle configuration the type of deformation resulting for a finite-sized system. Considering an elastic sphere, we here present such an approach. The method is in principle exact, at least within the framework of linear elasticity theory and for large enough interparticle distances. Different particle arrangements are considered. We find, for instance, that regular simple cubic configurations show elongation of the sphere along the magnetization if oriented along a face or space diagonal of the cubic unit cell. Contrariwise, with the magnetization along the edge of the cubic unit cell, they contract. The opposite is true in this geometry for body- and face-centered configurations. Remarkably, for the latter configurations but the magnetization along a face or space diagonal of the unit cell, contraction was observed to revert to expansion with decreasing Poisson ratio of the elastic material. Randomized configurations were considered as well. They show a tendency of elongating the sphere along the magnetization, which is more pronounced for compressible systems. Our results can be tested against actual experiments for spherical samples. Moreover, our approach shall support the search of optimal particle distributions for a maximized effect of actuation. 
 
\end{abstract}

\maketitle

\section{Introduction}

Magnetic gels and elastomers, also referred to as magnetorheological elastomers, magnetosensitive elastomers, ferrogels, or differently, are magnetoelastic hybrid composite materials of magnetic or magnetizable colloidal particles embedded in a soft polymer-based elastic matrix \cite{filipcsei2007magnetic,menzel2015tuned, odenbach2016microstructure,lopez2016mechanics,weeber2018polymer}. Many of their outstanding properties arise because they can be addressed by external magnetic fields. Through these fields, the magnetic interactions between the particles are affected, which presses or rotates the embedded particles against the surrounding elastic environment. As a consequence, the overall properties of the material are altered. For instance, in this way the mechanical stiffness can be tuned and adjusted to a certain amount as requested \cite{jolly1996magnetoviscoelastic,jolly1996model, jarkova2003hydrodynamics,filipcsei2007magnetic, stepanov2007effect,bose2009magnetorheological,chertovich2010new, wood2011modeling,ivaneyko2012effects,evans2012highly,han2013field, borin2013tuning,chiba2013wide,pessot2014structural,menzel2015tuned, sorokin2015hysteresis,pessot2016dynamic,volkova2017motion, oguro2017magnetic,pessot2018tunable,watanabe2018effect}. Such induced switching, because of the involved restoring elastic forces, is typically reversible \cite{schumann2017situ}. 

Here, we concentrate on a different magnetically induced effect, namely, on overall shape changes resulting from the modified particle interactions. Corresponding externally and reversibly induced magnetostrictive behavior can be exploited to construct, for example, soft actuation devices \cite{zrinyi1996deformation,collin2003frozen,an2003actuating, filipcsei2007magnetic,zimmermann2007deformable,raikher2008shape, fuhrer2009crosslinking,bose2012soft,ilg2013stimuli, schmauch2017chained,hines2017soft}. Naturally, in this context it becomes crucial to know whether the employed system or sample will contract or elongate along an applied magnetic field (usually involving opposite behavior along the transverse directions because of the typical quasi-incompressibility of the materials). Our central focus in the present work is on the question of the kind of overall resulting shape changes. 

In experiments, commonly an elongation of the investigated materials along the axis of the applied magnetic field is observed \cite{bednarek1999giant,ginder2002magnetostrictive, guan2008magnetostrictive,diguet2009dipolar,diguet2010shape, maas2016experimental,han2015magnetostriction,borin2019stress, gollwitzer2008measuring, filipcsei2010magnetodeformation}. 
Yet, as has been revealed by many theoretical studies, the type of expected deformation strongly depends on the internal structural particle arrangement of the investigated systems. For instance, simple regular rectangular lattice structures with magnetic moments induced along the edges of the rectangular unit cells were found to contract along these edges \cite{ivaneyko2011magneto,ivaneyko2012effects,metsch2016numerical}. An extreme example of this kind is given by just a pair of magnetic particles that are driven towards each other by induced magnetic attraction \cite{biller2014modeling,biller2015mesoscopic,puljiz2018touching}. In contrast to that, regular body-centered cubic lattices are found to extend along the induced axis of magnetization when oriented along the edges of the cubic lattice cells \cite{ivaneyko2012effects}, as are two-dimensional systems containing hexagonal or initially wiggled chain-like structures 
\cite{metsch2016numerical}. 
Likewise, the rotation of embedded clusters can lead to an extension along the applied magnetic field \cite{stolbov2011modelling,gong2012full}. 

One significant problem in theoretical treatments is that simplifications and approximations are mostly unavoidable when characterizing complex materials of the considered kind. Consequently, the results need to be treated with special care. Studies based on assumptions of affine (homogeneous) deformations suppress internal degrees of freedom that may become important \cite{wood2011modeling,ivaneyko2011magneto,ivaneyko2012effects}. Basic dipole-spring models may serve to include such internal degrees of freedom to a certain amount \cite{annunziata2013hardening,pessot2014structural,pessot2016dynamic, weeber2017arxiv,pessot2018tunable,ivaneyko2018dynamic, goh2018dynamics}, however, it is difficult to fully comply with quasi-incompressibility in such approaches. The same applies to more microscopic simulation approaches resolving in a coarse-grained manner individual polymer chains \cite{weeber2012deformation,weeber2015ferrogels,weeber2019studying}. Finite-element simulations, although quantitatively very reliable \cite{gong2012full,han2013field,biller2014modeling, metsch2016numerical,kalina2016microscale}, may be limited at some point concerning the number of the considered embedded particles. So far, the majority of these simulation works seems to have been carried out for bulk systems. Moreover, statistical and scale-bridging procedures likewise and naturally contain certain types of approximations \cite{zubarev2012theory,ivaneyko2014mechanical,menzel2014bridging, zubarev2015effect,romeis2016elongated}. 

To avoid many of the problems involved in the approaches just summarized, we adhere to the following strategy. We consider elastic spherical systems, as realized in some experiments \cite{gollwitzer2008measuring, filipcsei2010magnetodeformation}, to explicitly include the role of the system boundaries. 
Moreover, we confine ourselves to linear elasticity, that is, only distortions of low amplitude are addressed \cite{landau1986theory,puljiz2016forces,puljiz2017forces,puljiz2019displacement}. For elastic spheres embedded under no-slip surface conditions in a homogeneous and infinitely extended elastic background, an analytical expression for the linearly elastic Green's function is available \cite{walpole2002elastic}. It describes the static elastic displacements that result in response to a constant force acting onto the elastic material at one point-like force center. 
We adapt this expression to pairwise forces acting inside a free-standing elastic sphere without any additional elastic background medium. Elastic distortions of the sphere, particularly along its surface, are then evaluated when many such force centers are present. 
Consequently, the force centers are identified with the embedded magnetic particles, subject to pairwise magnetic interaction forces between them. In this way, we analyze the resulting overall deformation of the whole free-standing sphere upon induced magnetization of various different contained regular and randomized particle distributions.

\section{System under investigation} \label{sec_sytems}

\begin{figure}
\includegraphics[width=.9\linewidth]{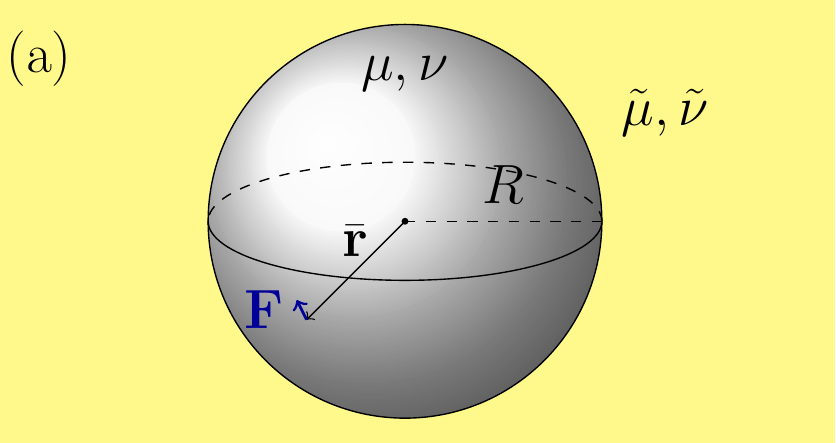} \\
\includegraphics[width=.9\linewidth]{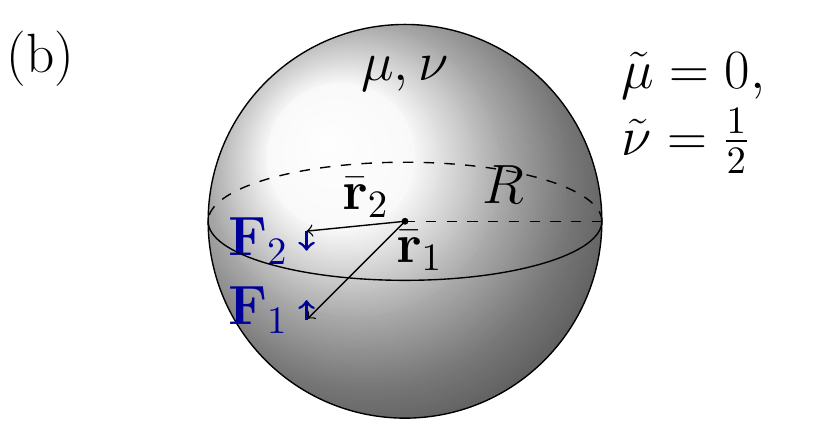}
\caption{Illustration of the system under investigation. (a) Walpole considered a deformable elastic sphere of elastic shear modulus $\mu$ and Poisson ratio $\nu$ embedded under no-slip surface conditions  in an infinitely extended elastic background medium of shear modulus $\tilde{\mu}$ and Poisson ratio $\tilde{\nu}$. He determined the corresponding Green's function, which quantifies the displacement field $\mathbf{u}(\mathbf{r})$ inside and outside the sphere resulting from a point-like force center acting on the inside \cite{walpole2002elastic}. (b) On this basis, we investigate the deformation of a free-standing sphere (for $\tilde{\mu}\rightarrow0$ and $\tilde{\nu}\rightarrow1/2$) that contains many point-like force centers exerting pairwise magnetic forces of vanishing global force on the sphere. In this case, terms in the original solution \cite{walpole2002elastic} that would diverge, connected to a net translation of the sphere, balance each other. Then, Walpole's solution can be adapted accordingly. Similar reasoning applies to net rotations of the sphere in response to net torques acting on it.}
\label{fig_system}
\end{figure}

As already indicated above, we confine ourselves to spherical elastic systems, strongly relying on previous work by Walpole \cite{walpole2002elastic}. In his study, a system as sketched in Fig.~\ref{fig_system}(a) was investigated. 
An elastic sphere of modulus $\mu$ and Poisson ratio $\nu$ is embedded in an infinitely extended elastic background of modulus $\tilde{\mu}$ and Poisson ratio $\tilde{\nu}$. The sphere of radius $R$ is centered around the origin. Both parts, the sphere and the surrounding elastic background, are linearly elastic, spatially homogeneous, and locally isotropic in their undeformed states. At the interface between the two parts, perfect bonding prevails, implying continuity of the corresponding displacement fields and traction vectors. 

In an impressive treatment of this problem, Walpole managed to derive the associated Green's function \cite{walpole2002elastic}. It solves the corresponding Navier--Cauchy equations \cite{cauchy1828exercises}
\begin{equation}\label{NC}
\mu\Delta \mathbf{u}(\mathbf{r}) + \frac{\mu}{1-2\nu}\nabla\nabla\cdot\mathbf{u}(\mathbf{r}) = -\mathbf{f}(\mathbf{r})
\end{equation}
inside the sphere ($r=|\mathbf{r}|<R$) and with the replacements $\mu\rightarrow\tilde{\mu}$ and $\nu\rightarrow\tilde{\nu}$ outside the sphere ($r>R$), respecting the described boundary conditions at the interface ($r=R$). In these equations, $\mathbf{u}(\mathbf{r})$ represents the displacement field, and the force density $\mathbf{f}(\mathbf{r})$ is specified as $\mathbf{f}(\mathbf{r})=\mathbf{F}\delta(\mathbf{r}-\mathbf{\bar{r}})$. That is, the Green's function $\mathbf{\underline{G}}(\mathbf{r},\mathbf{\bar{r}})$ provides the solution for the resulting displacement field $\mathbf{u}(\mathbf{r})$ in response to a static force $\mathbf{F}$. The force $\mathbf{F}$ acts on one point-like force center located at position $\mathbf{\bar{r}}$. It leads to the displacement field $\mathbf{u}(\mathbf{r})=\mathbf{\underline{G}}(\mathbf{r},\mathbf{\bar{r}})\cdot\mathbf{F}$. To evaluate $\mathbf{\underline{G}}(\mathbf{r},\mathbf{\bar{r}})$, we have implemented it numerically following the presentation in Ref.~\onlinecite{walpole2002elastic}. 

All force centers are located on the inside of the sphere ($\bar{r}=|\mathbf{\bar{r}}|<R$). It was already mentioned in Ref.~\onlinecite{walpole2002elastic} that the limit $\tilde{\mu}\rightarrow0$ leads to a divergence of $\mathbf{u}(\mathbf{r})$. From a physical point of view, this can be understood as follows. The limit $\tilde{\mu}\rightarrow0$ (and simultaneously $ \tilde{\nu} \rightarrow 1/2 $) implies the absence of an elastic background. Thus a free-standing sphere is considered, see Fig.~\ref{fig_system}(b). If a net force $\mathbf{F}$ acts on the sphere, there is no surrounding elastic background that would hold the sphere back from displacing.
Thus, an arbitrarily small but finite magnitude of the force $\mathbf{F}$ can displace the sphere by an infinite amount. 

Physical intuition then implies that the corresponding divergence for $\tilde{\mu}\rightarrow0$ should be lifted when pairwise reciprocal forces are considered. In this case, there are zero remaining net forces acting on the sphere. The sole effect of the forces is then to deform (or rotate) the sphere. 

We could demonstrate by analytical considerations that, in fact, the terms that lead to the described divergence lift each other for pairwise reciprocal forces.
The necessary condition for the forces is indeed satisfied for the pairwise magnetic forces considered in our present study. 
We thus drop the corresponding terms from the expressions listed in Ref.~\onlinecite{walpole2002elastic}. 

A similar divergence results, if the applied forces lead to a net torque on the sphere. Analogously, this situation leads to a divergence of the displacement field for $\tilde{\mu}\rightarrow0$. A free-standing sphere can be rotated by an infinite amount, if an arbitrarily small but finite net torque is applied to it.
Again, if the net torque vanishes, terms that would lead to the divergence lift each other, and we thus drop them from the solution in Ref.~\onlinecite{walpole2002elastic}.

We numerically implemented the corresponding expressions and confirmed their validity by comparison with extrapolated results obtained from the numerical implementation of Walpole's expressions \cite{walpole2002elastic} for decreasing $\tilde{\mu}$. To confirm the correctness of the latter, we tested that our implementation satisfies Eq.~(\ref{NC}) and the boundary conditions on the surface of the sphere. Moreover, we have tested that our numerical implementation of the Green's function reduces to the results from the well-known bulk Green's function \cite{teodosiu1982elastic} when we set $\mu=\tilde{\mu}$ and $ \nu=\tilde{\nu} $. Apart from that, for $\mu\neq0$ and $\tilde{\mu}\rightarrow\infty$, we considered force centers located close to the boundary of the sphere. In this case, the boundary can be approximated as a flat rigid wall. Our numerical results in this limit agree well with those of the Green's function calculated for a half-space filling elastic material bordered by a rigid no-slip boundary \cite{phan1983image,menzel2017force}. A similar solution exists for a half-space in the case of $ \mu \neq 0 $ and $ \tilde{\mu} \rightarrow 0 $ with $ \tilde{\nu} \rightarrow 1/2 $, i.e.\ a semi-infinite solid with a free boundary \cite{mindlin1936force}. Again, in the considered limit, the results of our numerical implementation match the results obtained from these analytical expressions.

As a result, the deformations induced by magnetic interactions of inclusions within a free-standing elastic sphere can be calculated. For simplicity, we concentrate on magnetic dipolar interactions between the inclusions \cite{klapp2005dipolar,weeber2012deformation}. All inclusions are assumed to be identical and to carry the same magnetic dipole moment $\mathbf{m}=m\mathbf{\hat{m}}$, with $m=|\mathbf{m}|$. Such a situation arises, for example, if a strong external magnetic field magnetizes all small identical spherical inclusions to saturation. Then the induced magnetic force acting on the $i$th inclusion, exerted by all other inclusions, reads \cite{jackson1962classical}
\begin{equation}
\mathbf{F}_i= -\, \frac{3 \mu_0 m^2}{4\pi} \sum_{\substack{j=1 \\ j \neq i}}^{N} 
		\frac{5 \mathbf{\hat{\bar{r}}}_{ij} \!\left( \mathbf{\hat{m}} \cdot \mathbf{\hat{\bar{r}}}_{ij}\right)^2 - \mathbf{\hat{\bar{r}}}_{ij} - 2 \mathbf{\hat{m}} \!\left( \mathbf{\hat{m}} \cdot \mathbf{\hat{\bar{r}}}_{ij} \right) }{\bar{r}_{ij}^4},
		\label{eq_magn_dipole}
\end{equation}
where $\mu_0$ denotes the magnetic vacuum permeability, $\mathbf{\bar{r}}_i$ sets the position of the $i$th inclusion, $\mathbf{\bar{r}}_{ij}=\mathbf{\bar{r}}_i-\mathbf{\bar{r}}_j=\bar{r}_{ij}\mathbf{\hat{\bar{r}}}_{ij}$ with $\bar{r}_{ij}=|\mathbf{{\bar{r}}}_{ij}|$, $i=1,...,N$, and $N$ fixes the number of inclusions.

In all that follows, we rescale lengths by the radius $R$ of the elastic sphere and measure forces in units of $\mu R^2$. Thermodynamic stability requires $\mu>0$ for the shear modulus of the sphere and $-1<\nu<1/2$ for the Poisson ratio \cite{landau1986theory}, while the limit $\nu \rightarrow 1/2$ characterizes an incompressible material. Negative Poisson ratios refer to so-called auxetic materials that, if stretched along one axis, expand to the lateral directions instead of contracting. 

Summarizing, the free-standing elastic sphere is distorted in response to the magnetic forces acting on the embedded magnetic inclusions. These inclusions are assumed to be spherical and of radius $ a=0.02 R $. To find the steady distorted state, we adhere to the following iterative numerical scheme. 

First, we estimate how a rigid sphere of radius $ a $ embedded inside the free-standing elastic sphere is displaced in response to a force acting on it, see the supplementary material (Sec.~\ref{suppl}). Analytical expressions fitted to the numerical estimates are obtained for these displacements, see again the supplementary material (Sec.~\ref{suppl}). Next, the forces on all other inclusions induce elastic distortions within the elastic sphere that add to the displacement of the considered inclusion. These additional contributions to the displacement are calculated from our modified version of Walpole's solution. To evaluate this mutual interaction, mediated by the elastic environment, the inclusions are treated as point-like, assuming them to be sufficiently far apart from each other. Then, after having calculated the new positions of all inclusions, we can evaluate the magnetic interactions between them anew. This in turn leads to different displacements, which again leads to adjusted forces, and so on. 
After multiple of these steps of iteration, we reach a steady state. Our goal is the final steady magnetic force distribution that can then be used in conjunction with our modified solution by Walpole to calculate the deformation on the surface of the elastic sphere within the framework of linear elasticity theory, once more treating the inclusions as point-like.

\section{Setting the remaining system parameters}\label{system-param}
Upon the mentioned rescaling, we obtain in Eq.~\eqref{eq_magn_dipole} a dimensionless force coefficient of $ 3 \mu_0 m^2/ 4\pi \mu R^6 $. To set its value in our subsequent evaluations in agreement with possible experimental realizations, we consider nickel or iron oxide (\chem{Fe_3 O_4}) as the material for the magnetic inclusions. For nickel, the literature, for instance, lists a saturation magnetization of approximately \SI{55.1}{\J\per\tesla\per\kg} \cite{crangle1971magnetization} or, by using the density of nickel of approximately \SI{8.908}{\g\per\centi\meter\cubed} \cite{nickel}, $ M_S \approx \SI{490.8}{\kJ\per\tesla\per\meter\cubed} (=\SI{490.8}{\kilo\ampere\per\meter}) $. For iron oxide, we use a saturation magnetization of approximately \SI{100}{\J\per\tesla\per\kg} \cite{cornell2003iron}, corresponding via a density of approximately \SI{5.18}{\g\per\centi\meter\cubed} \cite{cornell2003iron} to $ M_S \approx \SI{518}{\kJ\per\tesla\per\meter\cubed} (=\SI{518}{\kilo\ampere\per\meter})$. In both cases, we assume a shear modulus of \SI{1.67}{\kilo\pascal}. A further parameter is the radius of the magnetic inclusions $ a $ which we choose as $ a = 0.02 R $. Furthermore, we assume that the elastic material does not influence the magnetic interactions. These choices lead to a dimensionless force coefficient $ 3\mu_0 m^2/4\pi\mu R^6 $ of approximately $ 4.9 \times 10^{-8} $ for nickel or $ 5.4 \times 10^{-8} $ for iron oxide, respectively. The latter value is used in all our evaluations unless noted otherwise.

\section{Basic illustrative examples}\label{sec_illustration}

\begin{figure}[!t]
	\includegraphics[width=8.5cm]{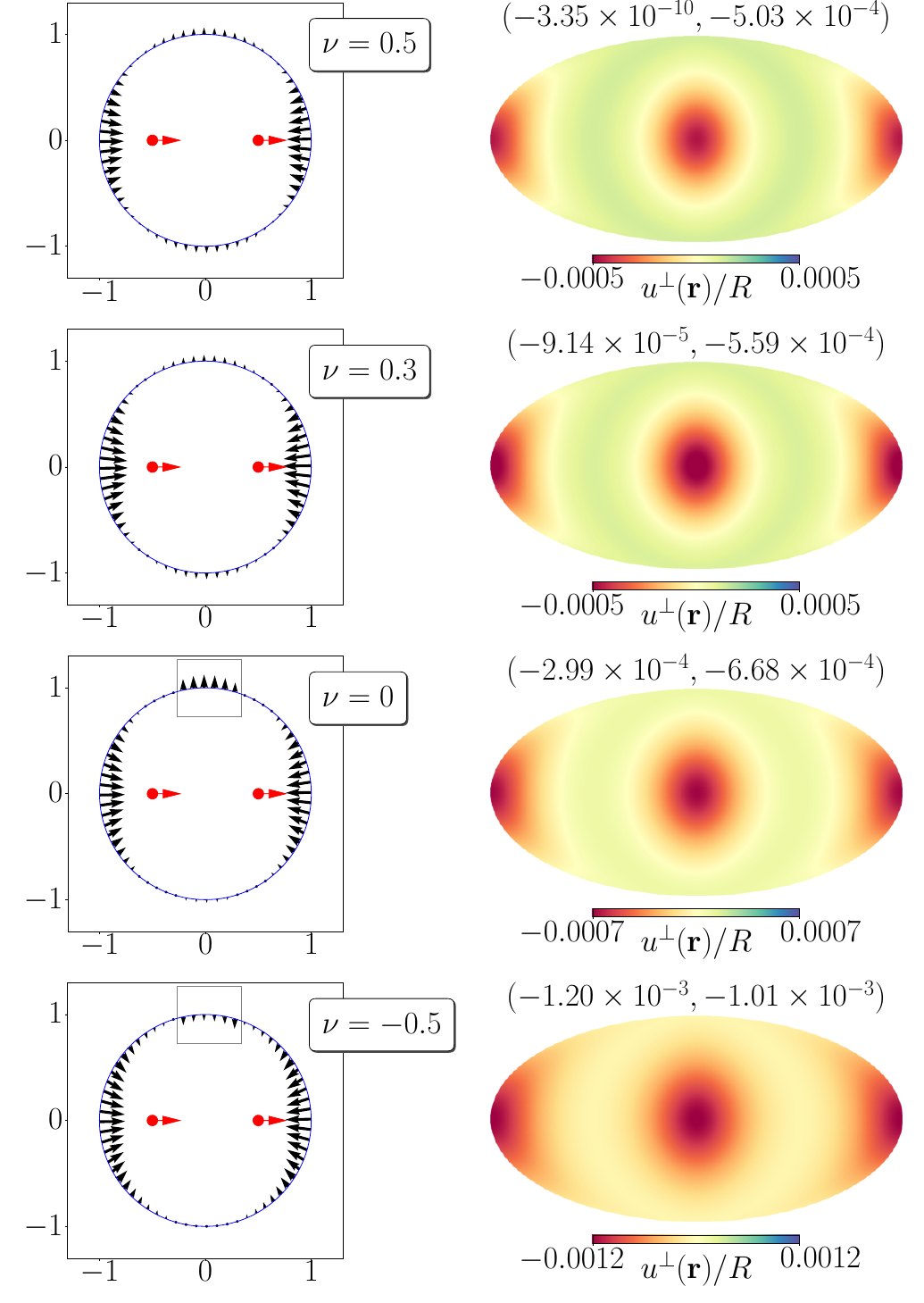}
\caption{Deformation of a sphere with two mutually attracting magnetic dipoles aligned along the horizontal symmetry axis, each at a distance of $0.5R$ from the center, and for $3\mu_0 m^2/4\pi\mu R^6=0.001$. The left column shows cross-sectional cuts through the sphere containing the symmetry axis. We mark the initial positions of the inclusions by red dots and the direction of the magnetic moments by red arrows. The surface displacements, indicated by black arrows, are enhanced by a factor of 500 for $ \nu > 0 $, 350 for $ \nu=0 $, and 200 for $ \nu < 0 $, respectively. Inside the gray frames, we enhanced the displacement arrows by an additional factor of 4. On the right-hand side, we illustrate the induced displacements on the surface of the sphere by so-called Mollweide projection plots. Outward displacements are marked in green / blue, inward displacements in orange / red, with the numbers on the scale bars encoding the radial displacements. All Mollweide plots were arranged so that the magnetic moments point outwards towards the reader from the centers of the plots and inwards on the left and right ends. Furthermore, all plots indicate a contraction of the sphere along the symmetry axis. For positive Poisson ratio $\nu$ the sphere expands along the lateral directions. In contrast to that, for the auxetic case of $\nu=-0.5$, a contraction along all directions is observed. The plots on the right-hand side were generated using the HEALPix package \cite{HEALPix}. Moreover, the pairs of numbers on each plot indicate the coefficients $(u^{\bot}_{00},u^{\bot}_{20})$ of an expansion into spherical harmonics of the radial outward displacement in units of $ R $, associated with the overall change in volume and relative elongation along the axis of magnetization, respectively.}
\label{fig_illustration1}
\end{figure}

\begin{figure}
	\includegraphics[width=8.5cm]{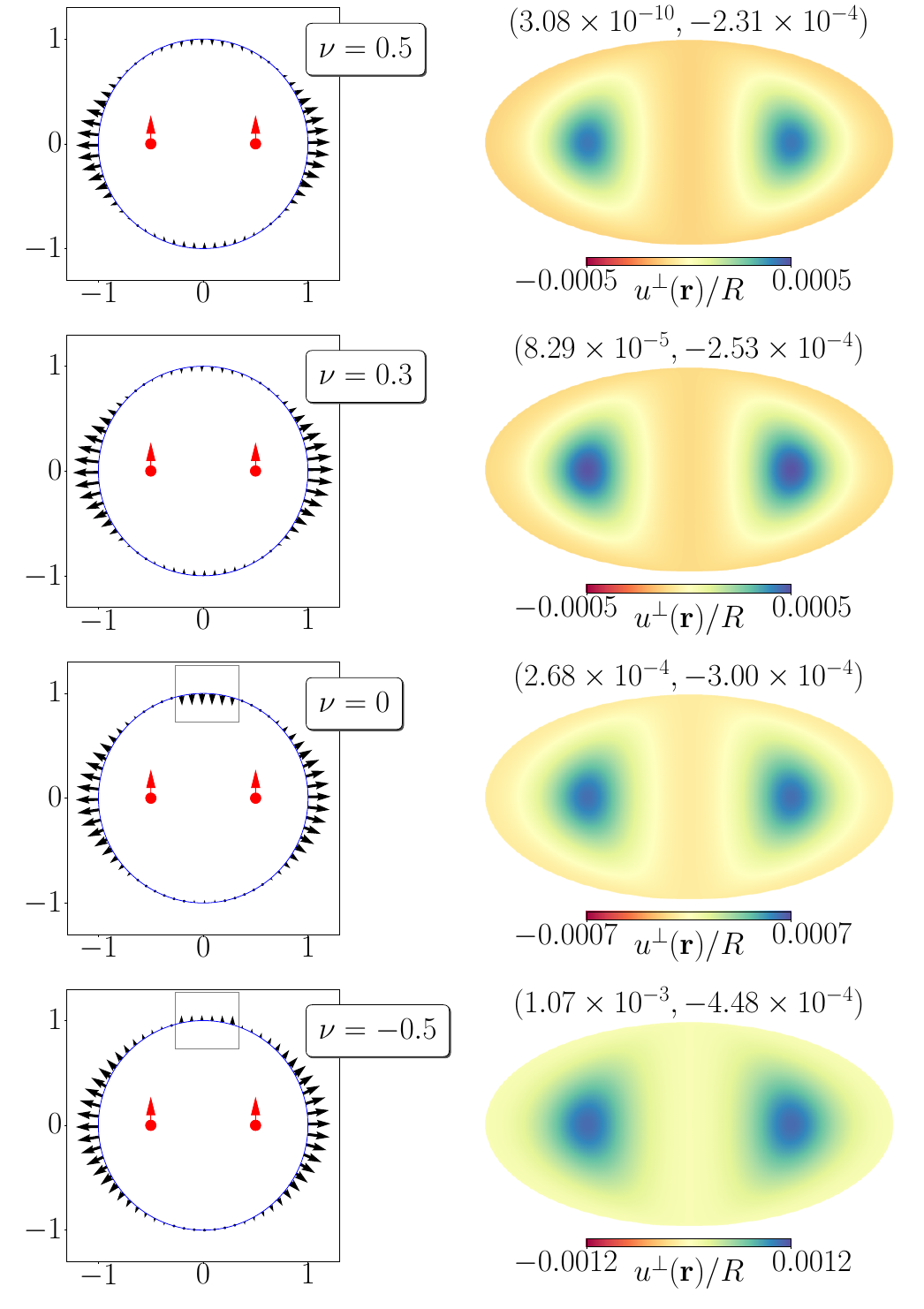}
	\caption{Same as in Fig.~\ref{fig_illustration1}, but now with the magnetic dipole moments located in a mutually repulsive configuration on the horizontal axis running through the center of the sphere and for $3\mu_0 m^2/4\pi\mu R^6=0.002$. The corresponding plots on the right-hand side were generated using  the HEALPix package \cite{HEALPix}. As in all our plots using the Mollweide projection, the magnetization vector points toward the reader at the center of each plot. The horizontal extension in the plots on the left-hand side due to the repulsion between the two dipoles is clearly visible by the dark blue spot in the Mollweide projections on the right-hand side. In the auxetic case of $ \nu = -0.5 $, an expansion along all directions is observed.}
\label{fig_illustration2}
\end{figure}

As a first step, we illustrate the formalism using two basic example situations as depicted in Figs.~\ref{fig_illustration1} and \ref{fig_illustration2}. Since here only two magnetic inclusions are considered, we use much higher dimensionless force coefficients than introduced in Sec.~\ref{system-param} to still produce visible displacements. First, in Fig.~\ref{fig_illustration1}, two mutually attractive magnetic dipole moments are induced on the horizontal symmetry axis running through the center of the sphere. As expected, the sphere in response to these induced forces contracts along the horizontal axis. For positive Poisson ratio, $\nu>0$, this contraction leads to an expansion in the lateral directions. This effect is most pronounced for an incompressible elastic sphere, i.e., for $\nu=0.5$, see the first row in Fig.~\ref{fig_illustration1}. In contrast to that, the negative Poisson ratio $\nu=-0.5$ reverses this secondary response, see the bottom row in Fig.~\ref{fig_illustration1}. 

A repulsive magnetic interaction between two inclusions is considered in Fig.~\ref{fig_illustration2}. Again, the magnetized particles are located on a horizontal axis running through the center of the sphere. In this case, the magnetic dipoles point into the vertical direction. As expected, the sphere now expands along the horizontal axis. Moreover, the sphere contracts along the vertical axis, except for the depicted case of $\nu=-0.5$, in which it expands along all directions. 

The right columns in both Figs.~\ref{fig_illustration1} and \ref{fig_illustration2} show how we illustrate our results in the following. To display the surface deformation, the resulting displacement field is evaluated on 49152 surface points. Then, the surface of the sphere is slit open and bent into the plane in a so-called Mollweide projection \cite{Mollweide}. By the color code we mark whether the surface is pushed towards the outside (green / blue) or pulled towards the inside (orange / red) of the sphere. Furthermore, the spheres are always rotated so that the magnetic moments point toward the reader in the center of the plots. We used the HEALPix package \cite{HEALPix} \!\!\footnote{http://healpix.sourceforge.net} to generate these plots. 

To obtain a more quantitative measure for the overall elongation or contraction of the sphere as well as for the overall change of volume, we proceed along the following lines. We determine for each of the $49152$ surface points the radial outward component $u^{\bot}(\mathbf{r}) = \mathbf{u}(\mathbf{r})\cdot \mathbf{\hat{r}}$ of the resulting displacement field $\mathbf{u}(\mathbf{r})$. Then, again using the HEALPix package \cite{HEALPix}, we expand $u^{\bot}(\mathbf{r}) $ into spherical harmonics \cite{jackson1962classical}. The coefficient $u^{\bot}_{00}$ of the spherical harmonic $Y_{00}=\sqrt{1/4\pi}$ indicates an overall expansion of the sphere (increased volume) for $u^{\bot}_{00}>0$ and an overall contraction (decreased volume) for $u^{\bot}_{00}<0$. Similarly, we determine the expansion coefficient $u^{\bot}_{20}$ for the spherical harmonic $Y_{20}=\sqrt{5/16\pi}\left(3\cos^2\!\theta-1\right)$, with $\theta$ denoting the angle of the center-to-surface vector on the sphere with respect to the magnetization direction. This coefficient, for $u^{\bot}_{20}>0$, indicates an elongation of the sphere along the axis of magnetization, relative to its transverse deformation. For $u^{\bot}_{20}<0$, contraction along the axis of magnetization, relative to the transverse deformation, occurs. 

Values of the corresponding pairs $(u^{\bot}_{00},u^{\bot}_{20})$ are indicated on the plots of Figs.~\ref{fig_illustration1} and \ref{fig_illustration2}. As expected, $u^{\bot}_{20}<0$ in Fig.~\ref{fig_illustration1} as well as in Fig.~\ref{fig_illustration2} due to the relative contraction along the direction of magnetization in both cases. Moreover, $ u^{\bot}_{00}<0 $ in Fig.~\ref{fig_illustration1} and $ u^{\bot}_{00}>0 $ in Fig.~\ref{fig_illustration2} (except for $ \nu=0.5 $) due to the mutual attraction and repulsion, respectively. The absolute values of $ u^{\bot}_{00} $ strongly increase for decreasing $ \nu $ as the sphere gets more compressible. Not shown are the displacements of the inclusions obtained from our iterative numerical procedure. These displacements are in each situation pointing into the directions of the forces, i.e.\ towards and away from the center in the attractive and repulsive case, respectively.

\clearpage
\onecolumngrid
\begin{figure*}[t]
	\includegraphics[width=.8\linewidth, trim={0.25cm 0.2cm 0.33cm 0cm}, clip]{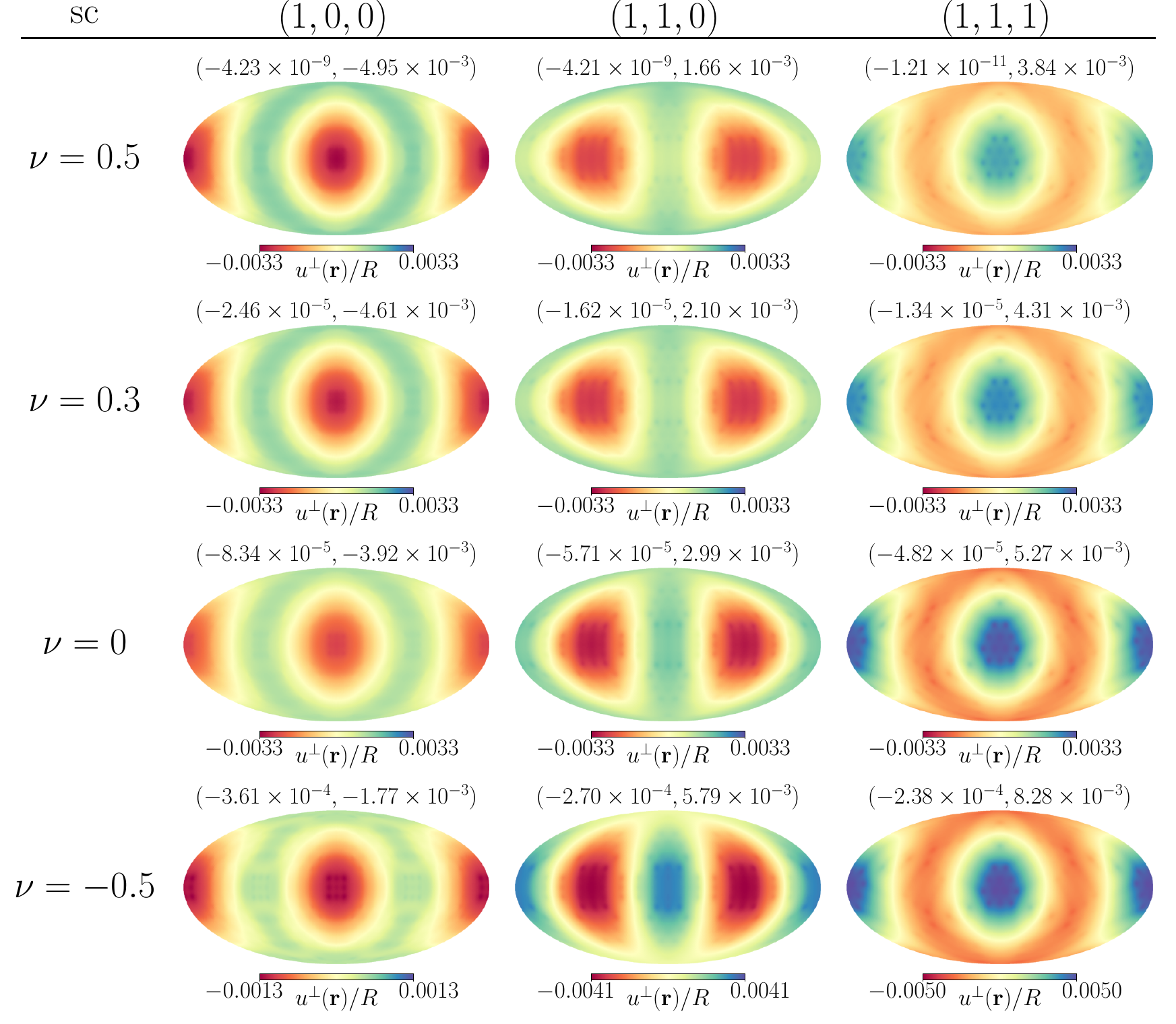}
	\caption{Mollweide projections and expansion coefficients $ (u^{\bot}_{00},u^{\bot}_{20}) $ in units of R \cite{HEALPix} for a simple cubic (sc) lattice structure of the embedded particle configuration, for $3\mu_0 m^2/4\pi\mu R^6= 5.4 \times 10^{-8}$. Again, the magnetization points towards the reader at the center of each plot.}
	\label{fig_sc}
\end{figure*}
\twocolumngrid 
\section{Many-particle configurations}
To now turn towards the situation in small model systems of magnetic gels and elastomers, we address structures of approximately $N=1000$ magnetic force centers distributed throughout the sphere. First, the effects of regular lattice configurations are analyzed for different orientations of the magnetization with respect to the corresponding unit cells. To position the force centers (``particles'') inside the sphere, we cut from a bulk-filling lattice structure all particles that are located on the inside of the sphere with a minimal distance of $3 a = 0.06 R$ from the surface. Afterwards, we also briefly illustrate situations of randomized particle distributions. As outlined at the end of Sec.~\ref{system-param}, we from now on set $3\mu_0 m^2/4\pi\mu R^6= 5.4 \times 10^{-8}$  throughout.

\subsection{Simple cubic lattice structure}\label{sec_sc}

We started with a configuration cut from a regular simple cubic (sc) lattice. As a lattice constant we chose $ 0.15 R $, which yields 1021 particles inside the elastic sphere. Three different orientations of the magnetization direction were probed, namely along the $(1,0,0)$ direction (edge of the unit cell), $(1,1,0)$ direction (face diagonal of the unit cell), and $(1,1,1)$ direction (space diagonal of the unit cell). See the left, center, and right column of plots in Fig.~\ref{fig_sc}, respectively. \\
First, we observe a contraction of the sphere along the magnetization axis when it is directed along the edge of the unit cell (left column of plots) in qualitative agreement with Ref.~\onlinecite{ivaneyko2012effects}. This can be understood already by considering the interactions within a pair of nearest neighbors. The corresponding dipole--dipole interactions are attractive along the magnetization direction and repulsive perpendicular to it. Second, in both other cases of orienting the magnetization, we find an expansion along the direction of the magnetization, which is more pronounced and uniaxial for the magnetization along the space diagonal of the unit cell (right column of plots). For the magnetization along the face diagonal (center column of plots), we observe contraction along the perpendicular face diagonal in the $ xy $-plane, i.e.\ in the $ (1,-1,0) $ or $ (-1,1,0) $ direction. Indicated by $ u^{\bot}_{00} $, we find that the volume of the sphere is shrinking in every case. 
\subsection{Rectangular lattice structure}
Next, we broke the cubic symmetry along one direction by turning to a rectangular lattice configuration. For this purpose, we considered a unit cell within which the lattice constant along one direction is 70\% of the lattice constants along the perpendicular directions. The latter lattice spacings were chosen as $ 0.171 R $, leading to 999 inclusions. We imposed the magnetization along the axis of smaller lattice constant as well as along one of the other axes. These cases are referred to by $ (1,0,0) $ and $ (0,1,0) $, respectively. \\
\begin{figure}[!t]
	\includegraphics[width=8.5cm, trim={0.24cm 0 2.3cm 0}, clip]{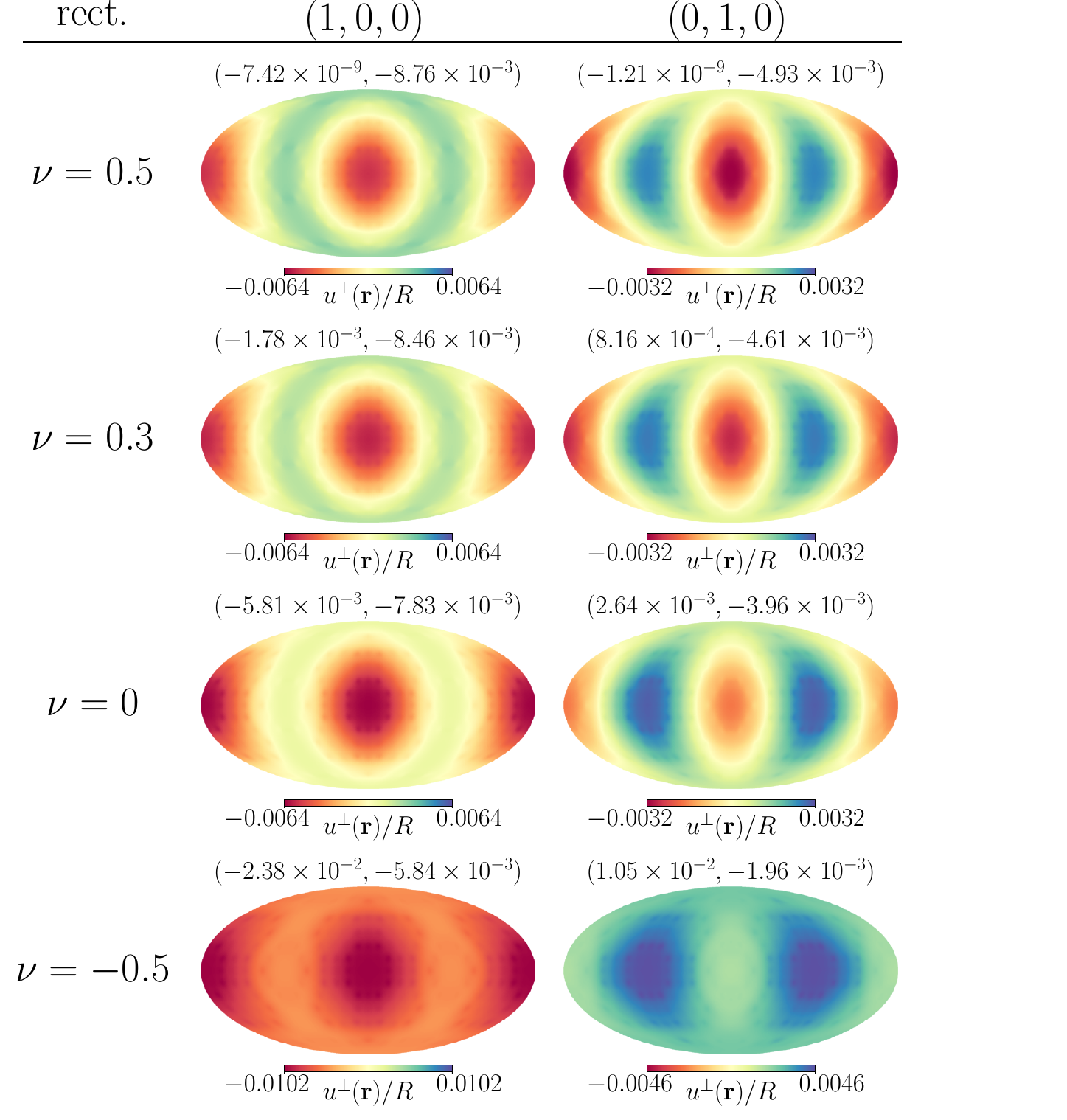}
	\caption{Same as in Fig.~\ref{fig_sc}, but for a rectangular (rect.) lattice structure featuring two identical edge lengths of the unit cell and the third edge length of 0.7 of that value, for $3\mu_0 m^2/4\pi\mu R^6= 5.4 \times 10^{-8}$. In the plots on the left-hand side, the magnetization is directed along the edge of smaller lattice constant, while it points along one of the edges of the larger lattice constant in the right-hand plots.}
	\label{fig_rect}
\end{figure}
Figure \ref{fig_rect} reveals a contraction of the sphere along the magnetization axis in both cases (except for $\nu=-0.5$). A relative contraction along this axis is observed in the spherical harmonic coefficients as $ u^{\bot}_{20} < 0$ in every case. In the auxetic situation ($ \nu = -0.5 $), a global contraction or expansion of the sphere is observed, respectively. We can understand this behavior from the interactions between the closest neighbors (along the axis of shorter lattice constant) which attract each other in the left-hand plots and repel each other in the right-hand plots in analogy to Sec.~\ref{sec_illustration}. Furthermore, we observe a four-fold symmetric deformation on the left-hand side due to the four-fold symmetry around the axis of magnetization, which is no longer present on the right-hand side. In the latter case, the symmetry axis parallel to the edges of shorter lattice constant runs through the center of the dark blue spots. Our results for the situation $ (1,0,0) $ and $ \nu > 0 $ can be compared to those of Ref.~\onlinecite{ivaneyko2011magneto} obtained for spatially homogeneous deformations of the elastic environment, showing qualitative agreement.

\clearpage
\onecolumngrid
\begin{figure*}[!t]
	\includegraphics[width=.8\linewidth, trim={0.25cm 0.2cm 0.33cm 0cm}, clip]{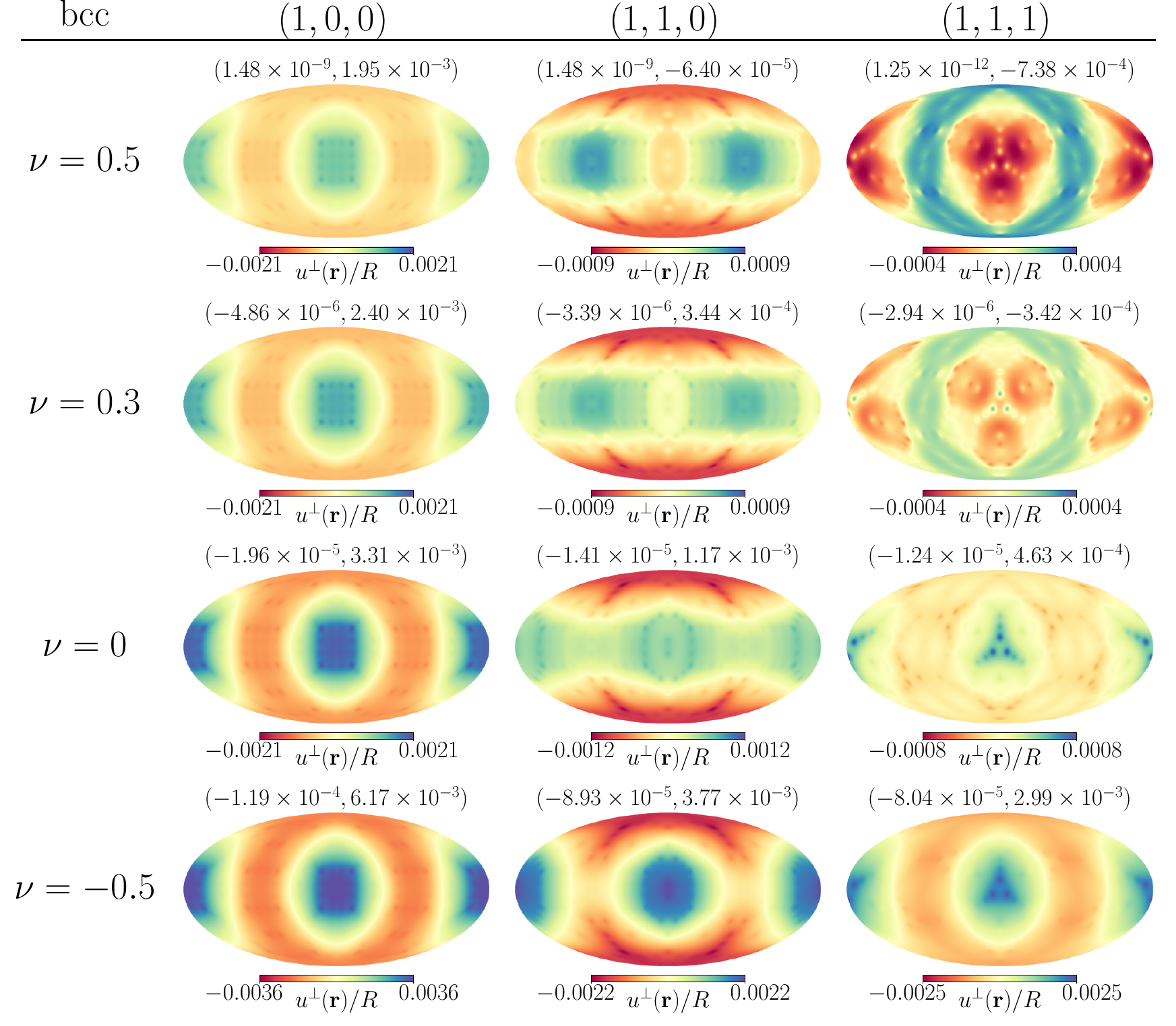}
	\caption{Same as in Fig.~\ref{fig_sc}, but for a body-centered cubic (bcc) lattice configuration, again setting $3\mu_0 m^2/4\pi\mu R^6= 5.4 \times 10^{-8}$.}
	\label{fig_bcc}
\end{figure*}
\twocolumngrid 

\FloatBarrier
\subsection{Body-centered cubic lattice structure}\label{sec_bcc}

Next, we address magnetostriction for body-centered cubic (bcc) lattice configurations. We probed the same magnetization directions $(1,0,0)$, $(1,1,0)$, and $(1,1,1)$ as for the sc lattice, see Sec.~\ref{sec_sc}. For comparison, these directions again refer to the cubic unit cell. A lattice constant for the cubic cell of $ 0.1885 R $ was used, which implies 1037 inclusions.
As a result, we observe a relative expansion along the magnetization axis for the $ (1,0,0) $-case, see Fig.~\ref{fig_bcc}. 
For the other two orientations of the magnetization, the opposite is true for $ \nu=0.5 $. Yet, for the auxetic spheres in the bottom row of Fig.~\ref{fig_bcc}, an expansion along the magnetization direction can be observed in all cases. Thus, for the $ (1,1,0) $ and $ (1,1,1) $ orientations, the response switches from contraction to expansion along the magnetization with decreasing $ \nu $.

\clearpage
\onecolumngrid
\begin{figure*}[!t]
	\includegraphics[width=.8\linewidth, trim={0.25cm 0.2cm 0.33cm 0cm}, clip]{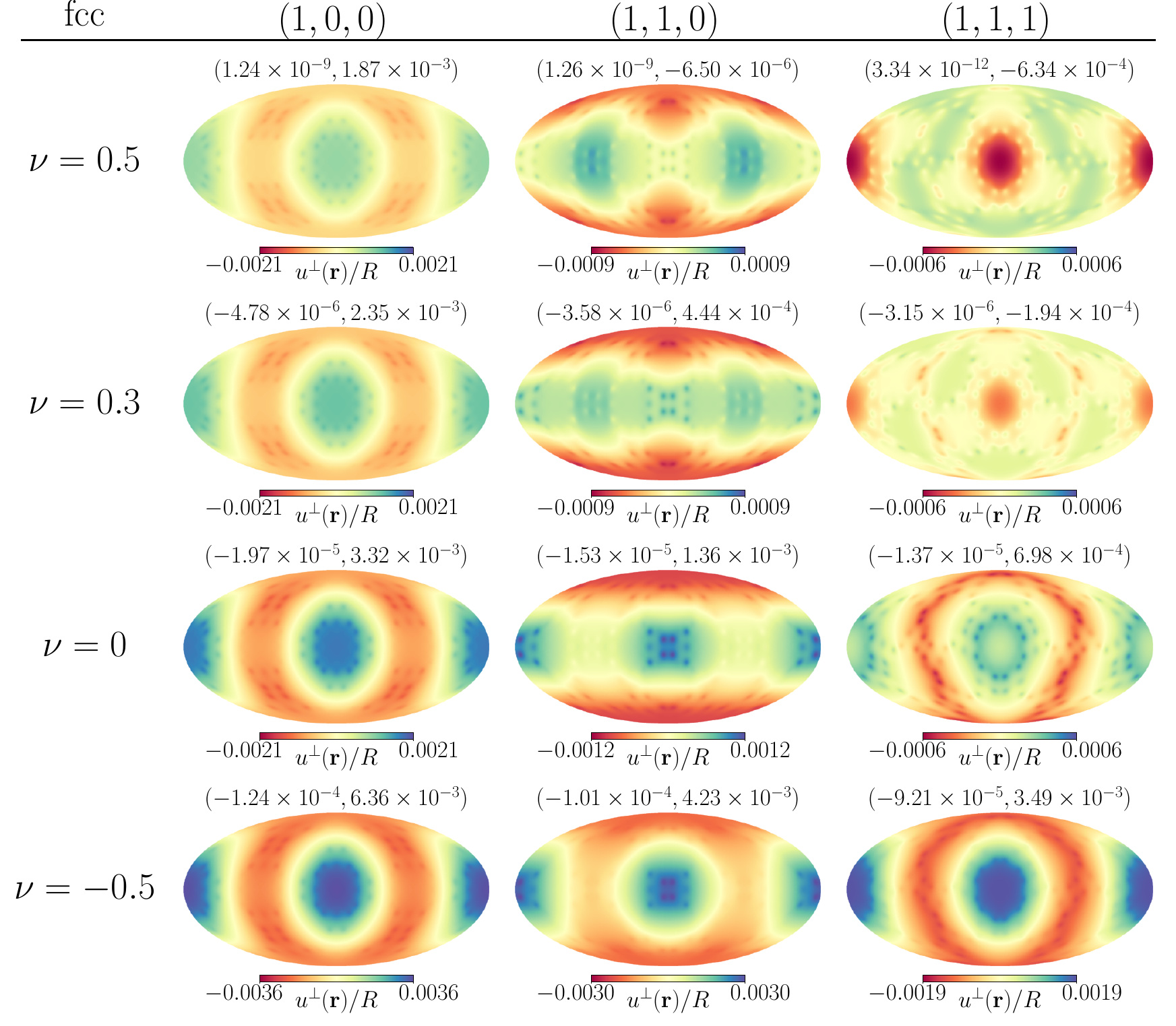}
	\caption{Same as in Fig.~\ref{fig_sc}, but for a face-centered cubic (fcc) lattice configuration, again setting $3\mu_0 m^2/4\pi\mu R^6= 5.4 \times 10^{-8}$.}
\label{fig_fcc}
\end{figure*}
\twocolumngrid 

\subsection{Face-centered cubic lattice structure}\label{sec_fcc}

Likewise, for face-centered cubic (fcc) lattice configurations, we label the orientations of the magnetization direction as $(1,0,0)$, $(1,1,0)$, and $(1,1,1)$. We used a lattice constant for the cubic cell of $ 0.236 R $, which implies 1055 inclusions.\\
In the case of this lattice structure, we find a strong resemblance to the previous situation of a bcc lattice, see Sec.~\ref{sec_bcc}, for all three magnetization directions that were probed.
Corresponding results are summarized in Fig.~\ref{fig_fcc}.
\subsection{Randomized configurations}

Finally, we also considered the response of less ordered particle configurations. For this purpose, the point-like particles were inserted at random into the sphere.
During the process, we impose a minimum distance of $6 a = 0.12 R$ of the particles from each other.

\begin{figure}[!t]
	\includegraphics[width=8.5cm, trim={0.24cm 0 2.3cm 0}, clip]{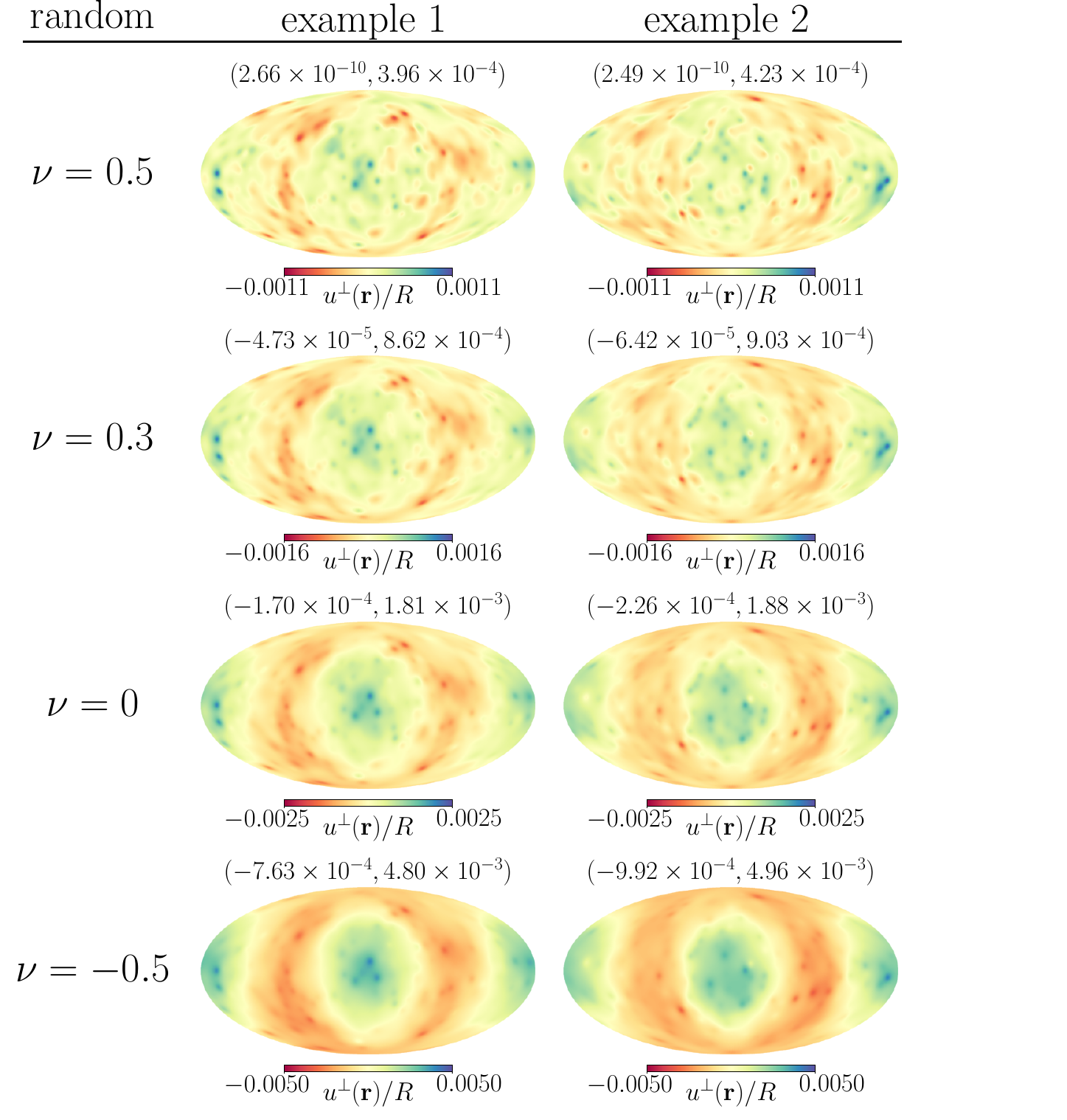}
	\caption{Same as in Fig.~\ref{fig_sc}, but for two randomized configurations, again setting $3\mu_0 m^2/4\pi\mu R^6= 5.4 \times 10^{-8}$.}
	\label{fig_rand}
\end{figure}

When aligned magnetic dipoles are generated on the particles, mutual magnetic interactions between them lead to deformations of the elastic sphere also for these more irregular configurations. Due to the disordered particle arrangement underneath the surface of the sphere, by eye the surface deformation appears quite irregular in most cases, see Fig.~\ref{fig_rand}. 
Generally, the observed irregularities in the shape of the deformed surface are in line with previous experimental observations and simulations \cite{stolbov2011modelling,gong2012full}. 

To obtain a definite measure for the overall volume changes and elongation or contraction along the axis of magnetization, we again determined the coefficients $u^{\bot}_{00}$ and $u^{\bot}_{20}$, respectively. As demonstrated for two randomly selected example realizations in Fig.~\ref{fig_rand}, the displacements generally tend to become more pronounced for decreasing $ \nu $. Our plots indicate an elongation along the magnetization axis, similarly to the situation for the fcc lattice magnetized along the $ (1,0,0) $ axis, see Fig.~\ref{fig_fcc}. To search for more general trends in the behavior of the values $u^{\bot}_{00}$ and $u^{\bot}_{20}$, in total 50 different realizations of randomized configurations were generated and evaluated.

\begin{figure}[!t]
	\includegraphics[width=8.5cm, trim={0 0.35cm 0 0.35cm}, clip]{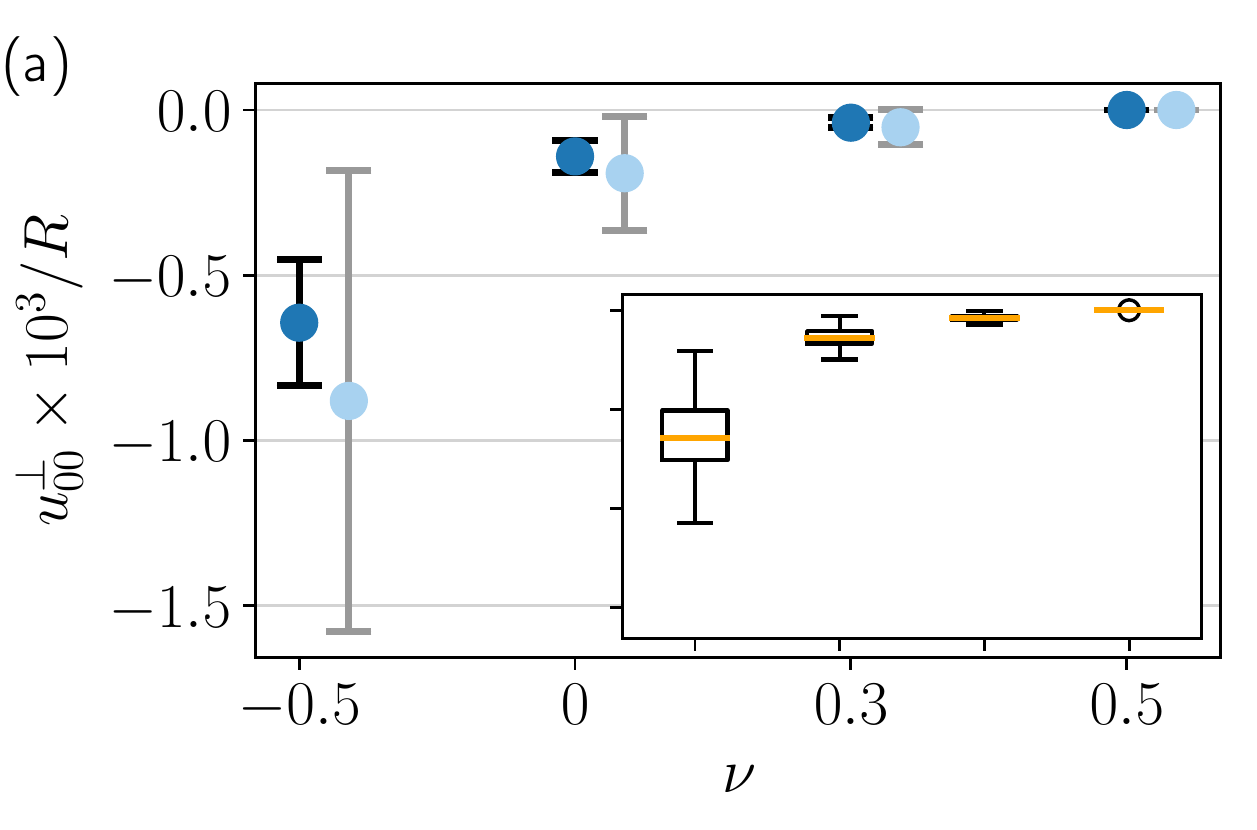}
	\includegraphics[width=8.5cm, trim={0 0.35cm 0 0.35cm}, clip]{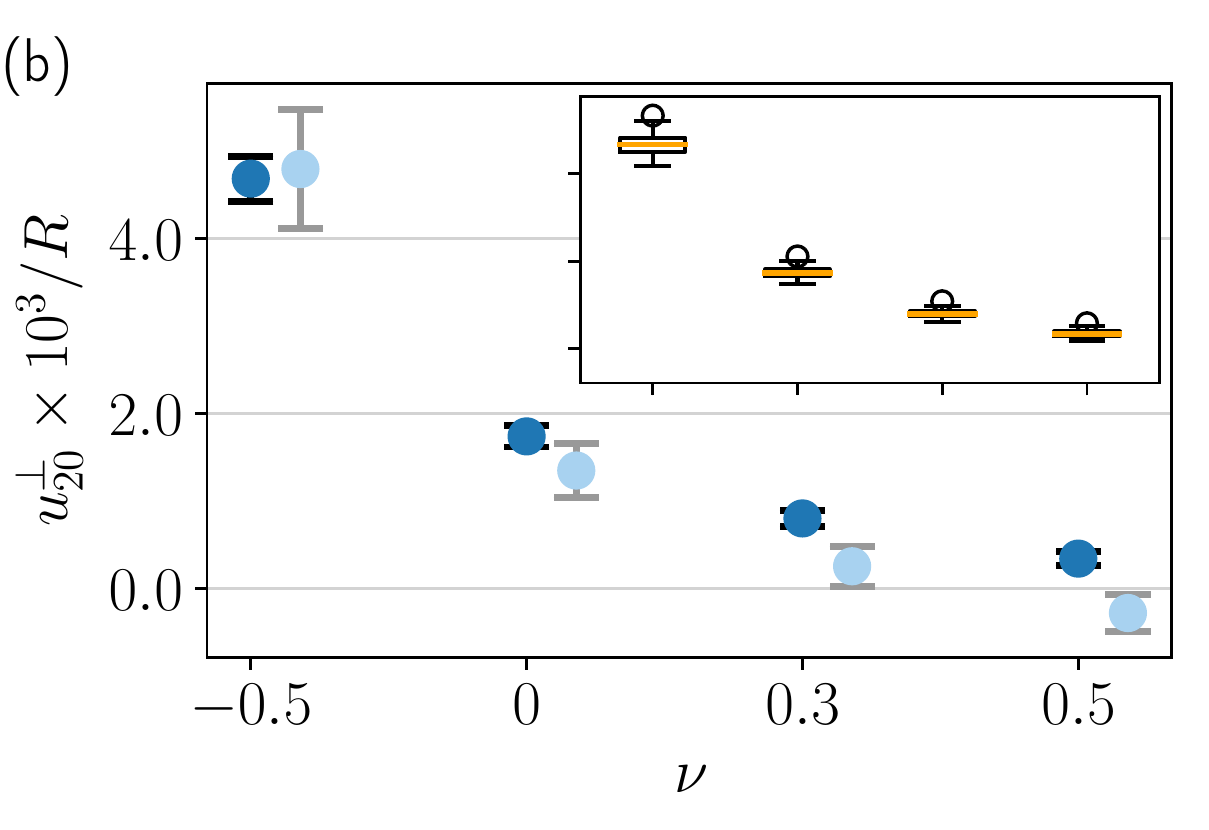}
	\caption{(a) Volume expansion ($u^{\bot}_{00}$) and (b) relative elongation along the magnetization axis ($u^{\bot}_{20}$) of the elastic sphere when randomized particle configurations within the sphere are magnetized as $3\mu_0 m^2/4\pi\mu R^6= 5.4 \times 10^{-8}$. We depict the corresponding mean values by darker (blue) dots and standard deviations by dark error bars obtained from 50 realizations of the particle distributions for the four studied values of the Poisson ratio (as marked on a nonlinear scale on the abscissae). Insets show corresponding boxplots, using the same scaling of the axes. The medians are plotted as (orange) horizontal lines. Boxes indicate the middle 50\% of obtained values, while the whiskers mark the highest and lowest values except for outliers that are shown as circles. Outliers are defined as points for which the distance to the end of the box is larger than $ 1.5 $ times the box height. Lighter symbols in the main plots were obtained for comparison for the same values of the Poisson ratio from three-fold mirror symmetric but otherwise randomized configurations.}
	\label{fig_rand_ulm}
\end{figure}

\begin{table}
\begin{tabular}{|@{\hskip 0.2cm} p{0.8cm} | @{\hskip 0.3cm} p{1.5cm} p{1.5cm} p{1.5cm} p{1.5cm} | }
		\hline
		$ \nu $ & $ -0.5 $ & $ 0 $ & $ 0.3 $ & $ 0.5 $\\
		$u^{\bot}_{00}$ & $ -6 \times 10^{-4} $ & $ -1 \times 10^{-4} $ & $ -4 \times 10^{-5} $ & $ 3 \times 10^{-10} $ \\
		$u^{\bot}_{20}$ & $ 5 \times 10^{-3} $ & $ 2 \times 10^{-3} $ & $ 8 \times 10^{-4} $ & $ 3 \times 10^{-4} $ \\
		\hline
\end{tabular}
\caption{Rounded averaged values of the expansion coefficients $u^{\bot}_{00}$ and $u^{\bot}_{20}$, averaged over 50 randomized configurations.}
\label{mean_rand_ulm}
\end{table}

The averaged results are summarized in Fig.~\ref{fig_rand_ulm}. A trend of overall contraction of the whole sphere is identified for smaller Poisson ratios, see the darker data points for $u^{\bot}_{00}$ in Fig.~\ref{fig_rand_ulm}(a). For an incompressible elastic sphere, i.e.\ for $ \nu = 0.5 $, the volume remains unchanged, as expected. In fact, we observe a trend of relative elongation of the sphere along the direction of magnetization ($u^{\bot}_{20} > 0$), see the darker data points in Fig.~\ref{fig_rand_ulm}(b). The absolute values quantifying the degree of deformation increase with decreasing Poisson ratio $ \nu $.
For clarity, we also summarize the averaged values in Tab.~\ref{mean_rand_ulm}. Comparing with related experimental results \cite{gollwitzer2008measuring,filipcsei2010magnetodeformation}, we find qualitative agreement.

We remark that, in general, the randomized particle configurations exposed to an external magnetic field will experience a net torque that would induce a macroscopic rotation. Still, we were able to perform our calculations. In our approach, as explained before, we have excluded corresponding contributions.
For comparison, to generate systems of vanishing overall torque, we changed our procedure of initialization. The particles were now inserted at random into that eighth of the sphere of Cartesian coordinates of $x,y,z > 0$. Afterwards, the configuration was mirrored at the planes $x=0$, $y=0$, and $z=0$ so that the whole sphere is filled.
The resulting data points for the same values of the Poisson ratio $\nu$ are included in lighter color in Fig.~9 and confirm the previously inferred trends. As one significant deviation, for incompressible systems ($\nu=0.5$), we now observe a weak tendency of relative contraction instead of elongation along the axis of magnetization. We can understand this deviation illustratively. The procedure of mirroring introduces pairs of nearby particles that mutually attract each other along or repel each other perpendicularly to the axis of magnetization, supporting an overall relative contraction of the sphere. We have checked that the associated change in the overall behavior for $\nu=0.5$ is related to the finite size of the sphere. When we double the radius $R$ of the elastic sphere, keeping the concentration of contained particles constant, we found the mean value in Fig.~\ref{fig_rand_ulm}(b) for $\nu=0.5$ and for the three-fold mirror-symmetric configuration (lighter data point) to move towards positive numbers. It would be interesting to know whether global rotations were observed in the macroscopic experiments \cite{gollwitzer2008measuring,filipcsei2010magnetodeformation}.
\section{Conclusions}

In summary, we have adapted the Green's function derived by Walpole for an elastic sphere embedded in an infinitely extended elastic environment \cite{walpole2002elastic} to the case of a free-standing elastic sphere. As a result, we can in the linear regime of elasticity calculate the small-amplitude deformations of the sphere, if ensembles of point-like inclusions exert forces in the absence of global translations and rotations. 

Our approach has the advantage of explicitly including the boundaries of the system, incompressibility or auxetic behavior if required, and all internal degrees of freedom of the elastic matrix. It treats the matrix as an elastic continuum. In principle, the formalism becomes exact for linear small-amplitude distortions and low volume fractions of the inserted particles, that is, large interparticle distances when compared to the particle diameters. At the same time, the action of many force centers can be conveniently superimposed in this way.
We stress that this formalism naturally takes into account the spatial inhomogeneities of deformation that occur within the elastic material. 

Along these lines, we have addressed the deformation of spherical elastic example systems that contain different spatial arrangements of magnetized point-like inclusions. We started from the illustrative examples of two mutually attracting or repelling dipoles. Next, we addressed several different types of configurations of approximately one thousand inclusions. Interestingly, for simple cubic configurations, we observed contraction or extension along the magnetization direction, depending on whether the magnetization was along the edge or face / space diagonal of the unit cells. Body- and face-centered cubic configurations, in contrast to that, led to expansion along the magnetization direction when oriented along the edges of the unit cells. Remarkably, for magnetizations along the face and space diagonals of the latter two lattice types, the behavior switched from relative contraction to relative expansion along the magnetization direction with decreasing Poisson ratio. Moreover, simple rectangular and randomized configurations were addressed. 

Out of these examples, spheres containing randomized particle configurations probably represent the most relevant considered systems concerning actual experimental realizations using presently available tools of fabrication. These spheres tend to elongate along the axis of magnetization, which becomes more pronounced with decreasing Poisson ratio. Moreover, with decreasing Poisson ratio, they tend to decrease their volume upon magnetization. 
By construction, our approach is restricted to the linear regime of deformation. This means that only small-scale deformations of the elastic spheres can be described.

In reality, it is possible to generate spherical samples of magnetic gels and elastomers by curing the material in a spherical compartment \cite{gollwitzer2008measuring,filipcsei2010magnetodeformation}. Afterwards, the internal structure and overall magnetostriction induced by an external magnetic field can be reconstructed, for example, using x-ray micro-tomographic analyses \cite{gundermann2017statistical, schumann2017characterisation}. Small-scale spherical samples could likewise be generated using different methods like solvent evaporation \cite{hai2009preparation} or microfluidic methods \cite{chen2009microfluidic}. Another type of complex polymeric material, spherical samples of which were produced in the latter way successfully, are liquid-crystalline elastomers. They can likewise show induced deformations and actuation  \cite{ohm2009continuous,fleischmann2012one}. Magnetic gels and liquid-crystalline elastomers share several similarities in their overall stress-strain properties \cite{menzel2009nonlinear,
cremer2016superelastic}.
Potentially, also the induced distortions of spherical samples of magnetic gels generated by microfluidic production techniques \cite{kim2007fabrication,hwang2008microfluidic} reveal such related behavior. Possibly, our approach may further be helpful in characterizing the deformational response of biological cells containing embedded magnetic particles \cite{huang2002three}. 

\vspace*{0.15cm}
\section*{Supplementary material}
See the supplementary material (Sec.~\ref{suppl}) for further details on how we obtain approximate analytical expressions for the displacement of a finite-sized inclusion in our free-standing elastic sphere when the inclusion is subject to an applied net force.

\begin{acknowledgments}
Some of the results in this paper have been derived using the HEALPix package \cite{HEALPix}.
The authors thank the Deutsche Forschungsgemeinschaft for support of this work through the priority program SPP 1681, grant no.~ME 3571/3.
\end{acknowledgments}


\bibliography{lit_fischer}

\FloatBarrier
\cleardoublepage

\setcounter{figure}{0}
\renewcommand{\thefigure}{S\arabic{figure}}
\renewcommand{\theHfigure}{Supplement.\thefigure}

\setcounter{equation}{0}
\renewcommand{\theequation}{S\arabic{equation}}
\renewcommand{\theHequation}{Supplement.\theequation}
\section{Supplementary material} \label{suppl}

In this supplemental file, we describe how we calculate in our free-standing elastic sphere the displacement of each finite-sized inclusion when it is subject to a non-vanishing but sufficiently small force.

In our iterative numerical scheme, it is necessary to know during each step how much each magnetic inclusion is displaced directly in response to the force acting on it. This force, in our situation, results from the magnetic interactions with all other inclusions. Here, we assume the inclusions to be of rigid spherical shape of radius $ a = 0.02 R $, where $ R $ is the radius of the free-standing embedding elastic sphere.

We note that because of the spherical geometry of the overall elastic body in combination with the linearity of the underlying equations, we only need to consider two complementary situations: First, the force may act parallel to the positional vector of the inclusion within the sphere, that is along a radial axis. Second, the force may act perpendicular to this axis. In general, we can split any force into these two orthogonal components.

To take into account the finite size of the spherical inclusion, we distribute about $ 3 \times 10^6 $ point-like force centers approximately evenly on a spherical shell around the center of the inclusion (again using the HEALPix package \cite{HEALPix}). Then, we let a force of unit magnitude, divided by the number of force centers, act on each of these force centers. The resulting displacement field is evaluated on the spherical shell at 192 different positions, again distributed approximately evenly.

We calculate this displacement field using our modified version of Walpole's solution. At the end, we determine from the 192 positions of evaluation the average displacement. The latter is assigned as the displacement of the whole inclusion. (Corresponding standard deviations over the shell are calculated as well.)

In principle, we need to repeat this procedure for all positions along one (arbitrary) radial axis. Obviously, this is only possible for a finite number of center positions of the inclusion.  Therefore, we interpolate our results for intermediate positions. To achieve this, we use a fitting function for the resulting displacements of the form \\ \nolinebreak
\begin{equation}
u^d(\bar{r}) = \sum_{i=0}^{6} \frac{\alpha_i^d}{(1-\bar{r})^i} \; ,
\label{fit}
\end{equation}
where $ d \in \{\parallel, \bot \} $ marks displacement components in response to forces parallel or perpendicular to the radial axis, $ \bar{r} $ sets the distance of the center of the inclusion from the center of the elastic sphere, and $ \alpha_i^d $ are fit parameters.\\

\begin{figure*}[t]
	\includegraphics[width=.79\linewidth]{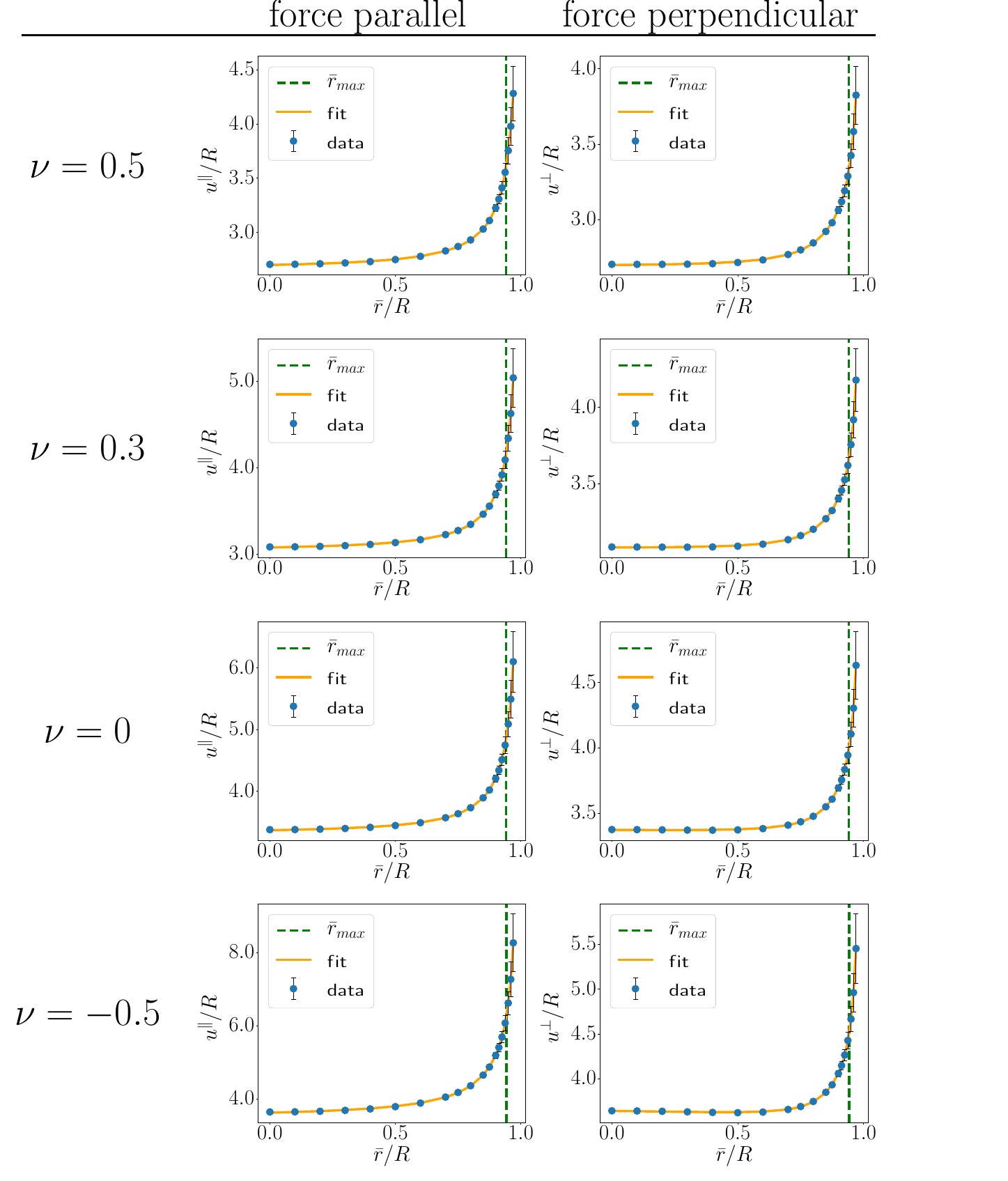}
	\caption{A spherical shell of radius $ a = 0.02 R $ is located at a distance $ \bar{r} $ from the center of a free-standing elastic sphere of radius $ R $ centered around the origin. As a measure for the induced displacements of a spherical inclusion of radius $ a $, we distribute point-like force centers on the shell. Each force center exerts a force of identical magnitude on the elastic body. Resulting displacements are evaluated at probe positions also located on the shell. We distinguish cases in which the force is applied parallel to the positional vector, i.e., along a radial axis (on the left-hand side), and cases in which the force is applied perpendicular to the positional vector (on the right-hand side). The (blue) dots represent mean values calculated from 192 probe positions distributed over each shell, together with corresponding standard deviations. Results for the fitting functions in Eq.~\eqref{fit} are indicated by solid (orange) lines. The vertical dashed (green) lines mark the maximal value of $ \bar{r} $ used in the evaluations reported in the main article. As in the main article, we distinguish four values of the Poisson ratio $ \nu $.}
	\label{fig_plots}
\end{figure*}

Figure \ref{fig_plots} demonstrates very good agreement between our fits using the functional form of Eq.~\eqref{fit} and our calculated data. Thus, we used the fitted functions to calculate the displacements of the inclusions during the iterative numerical loop as described in the main article. Naturally, the curves for the cases of displacement parallel and perpendicular to the radial axis coincide for $ \bar{r} = 0 $ for all considered values of the Poisson ratio $ \nu $. This is expected for symmetry reasons. In the limit $ \bar{r} / R \rightarrow 1 $, we compare the resulting displacements to corresponding displacements calculated via the solution by Mindlin \cite{mindlin1936force}. The latter refers to a homogeneous elastic half-space bounded by a free surface. We find good agreement for all four values of the Poisson ratio $ \nu $ and both force directions (data not shown). We note the increased standard deviation in Fig.~\ref{fig_plots} of the displacements on the shell of evaluation around the center of the inclusion for positions closer to the elastic surface. The points on that side located closer to the surface of the elastic sphere get displaced more than on the opposite side because there is less elastic material in the closer vicinity that needs to be dragged along (the elastic sphere ends within a finite distance). In our numerical evaluations associated with the results in the main article, we only used values $ \bar{r} < \bar{r}_{max} $, see Fig.~\ref{fig_plots}, for which these deviations are reasonably small.

\clearpage
\onecolumngrid

\end{document}